\documentclass[onecolumn]{aa} 
\usepackage{graphicx}
\usepackage{txfonts}
\begin{document}
   \title{A formation mechanism for the large plumes in the prominence}

   \author{Jincheng Wang\inst{1,2,3} 
   \and Xiaoli Yan\inst{1,3} 
   \and Zhike Xue\inst{1,3} 
   \and Liheng Yang\inst{1,3}
   \and Qiaoling Li\inst{4} 
   \and Hechao Chen\inst{5}
   \and Chun Xia\inst{6}
   \and Zhong Liu\inst{1,3}
    }
   \institute{Yunnan Observatories, Chinese Academy of Sciences, Kunming Yunnan 650216, PR China\\
         \email{wangjincheng@ynao.ac.cn}
         \and
             CAS Key Laboratory of Solar Activity, National Astronomical Observatories, Beijing 100012, PR China
            \and
            Center for Astronomical Mega-Science, Chinese Academy of Sciences, 20A Datun Road, Chaoyang District, Beijing, 100012, PR China
            \and
             Department of Physics, Yunnan University, Kunming 650091, People’s Republic of China
            \and
            School of Earth and Space Sciences, Peking University, Beijing 100871, People’s Republic of China
            \and
            School of Physics and Astronomy, Yunnan University, Kunming 650050, People’s Republic of China
             }

   \date{Received **** **, ****; accepted **** **, ****}
   
   \abstract
   {}
   {To understand the formation mechanism of large plumes in solar  prominences, we investigate the formation process of two such phenomena.}
   { We studied the dynamic and thermal properties of two large plumes using observations from New Vacuum Solar Telescope, the Solar Dynamic
Observatory, and the Solar Terrestrial Relations Observatory-Ahead. We employed the differential emission measures method to diagnose the thermodynamical nature of the bubble and plumes. We calculated the Doppler signals based on observations of H$\alpha$ blue and red wings.}
   {We find that two large plumes observed with high-resolution data are quite different from previously studied small-scale plumes. They are born at the top of a prominence bubble with a large projected area of 10-20 Mm$^2$. Before the occurrence of each large plume, the bubble expands and takes on a quasi-semicircular appearance. Meanwhile, the emission intensity of extreme-ultra-violet (EUV) bands increases in the bubble. A small-scale filament is found to erupt in the bubble during the second large plume. At the point at which the height of the bubble is comparable with half the width of the bubble, the bubble becomes unstable and generates the plumes. During the formation of plumes, two side edges of the top of the bubble, which are dominated by opposite Doppler signals, approach each other. The large plume then emerges and keeps rising up with a constant speed of about 13-15 km/s. These two large plumes have temperatures of $\sim$ 1.3 $\times$ $10^6$ Kelvin and densities of $\sim$ 2.0 $\times$ {10}$^9$\ {cm}$^{-3}$, two orders hotter and one order less dense than the typical prominence. We also find that the bubble is a hot, low-density volume instead of a void region beneath the cold and dense prominence.
}
    {These two large plumes are the result of the breakup of the prominence bubble triggered by an enhancement of thermal pressure; they separate from the bubble, most likely by magnetic reconnection.}
    
 \maketitle
 \section{Introduction}
 Solar prominences are some of the most spectacular structures in the Sun, located in the solar limb. They consist of dense, cool plasma suspended in the hot and tenuous corona,  the main potential source of solar storms. When they are observed in the solar disk, they appear as dark ribbon-like features composed of bundles of threads, also referred to as solar filaments \citep{mar98, mac10}. They often lie above magnetic polarity inversion lines (PILs) in the photosphere and are supported by sheared or twisted magnetic structures (e.g., sheared-arcade or flux rope) against solar gravity (e.g., \citealp{yang14,via15,yan15, wan17,che20}). High-resolution observations show that the on-disk filaments are mainly dominated by horizontal thin threads while the limb prominences are mainly dominated by vertical thin threads \citep{cha08,yan15b,li18}. Further, some horizontal threads or horizontal motions can be observed in some limb prominences \citep{ahn10,she15,yang18}. Some authors propose that the vertical thread structures in the prominences are a pile-up of dips in more or less horizontal magnetic field lines rather than real vertical magnetic structures \citep{sch10,dud12}. However, the relationship between these horizontal and vertical threads is still  controversial because of a lack of high-resolution stereoscopic observations.

Observations of the chromospheric spectral line (typically in H$\alpha$ and Ca II H lines) show that prominence bubbles are quasi-void regions beneath pre-existing prominences \citep{det08,ber08,ber11}. In extreme-ultraviolet (EUV) observations, bubbles are brighter than the main prominence and exhibit brightness  comparable to that of the surrounding corona \citep{she15,ber11,lab11}. Using the two-filter ratio technique, or differential emission measures (DEM) method, the temperature of bubble has been estimated to be on the order of $10^6$ Kelvin, which is much hotter than its overlying prominence ( $\sim$ $10^4$ Kelvin) and similar to the background corona \citep{she15,ber11,awa19}. The arcade fields of parasitic bipoles are thought to be the magnetic structures in the prominence bubbles \citep{dud12,gun14}. More recently, the boundaries of two prominence bubbles were found to correspond to the interface between the prominence barb and the underlying magnetic loops rooted nearby, and that the bubble can be formed around a filament barb seen in the solar disk \citep{guo21}. Some authors have suggested that the prominence bubbles are filled with hot and low-density plasma \citep{she15,ber11}, while others argue that the prominence bubbles are open windows that allow the passage of background coronal radiation \citep{dud12,gun14}. The nature of prominence bubbles remains controversial and merits further investigation.

High-resolution observations of prominences are revealing increasing detail in the prominences, including dark upflow plumes. The upflow plumes appear dark in the chromospheric lines first reported by \cite{ste73} and propagate initially from large bubbles and disperse into the prominences \citep{ber08,ber10,ber11}. \cite{det08} reported a rising bubble with a bright core traversing vertically through a prominence and suggested that the rising bubble is a void local system of stronger magnetic field with weak plasma pressure. Using observations of the Solar Optical Telescope (SOT) on board the Hinode satellite, \cite{ber10} investigated the kinematic characteristics of bubbles in three prominences, finding a mean ascent speed for plumes of up to 13 - 17 km/s and typical lifetimes of plumes ranging from 300s to 1000s. Using high-resolution observations of H$\alpha$ observed with the New Vacuum Solar Telescope,  \citet{xue21} found high levels of turbulence with large Doppler and nonthermal velocities widely occurring at the plume fronts. These dark upflow plumes are also an important source of quiescent prominence mass \citep{ber10}. Based on numerous observations and numerical simulations, authors proposed that dark upflow plumes are generated by the magnetic Rayleigh–Taylor instability (RTI) taking place at the boundary between the prominence and dark bubbles that rise into it \citep{ber11,ber10,hil12}. Kelvin–Helmholtz instability (KHI) at the boundary between the bubble and the prominence could also drive the occurrence of plumes \citep{ber17,awa19,xue21}. Based on these theories and some observational quantities, the plasma $\beta$ (ratio of gas pressure to magnetic pressure) was estimated in the prominence \citep{hil12b}. Therefore, the diverse properties of plumes and bubbles offer a unique toolkit with which to improve our understanding of solar prominences.

In this paper, we present high-resolution observations of two large plumes created by the breakup of a prominence bubble beneath a pre-existing prominence. These provide a rare opportunity to further our understanding of plume formation. By combining these observations with data from ground-based and space-based telescopes, we carry out a detailed study of the dynamic and thermal characteristics of plumes and the related bubble. The sections of this paper are organized as follows. The observations and methods are described in Section 2, the results are given in Section 3, and discussions and a summary are presented in Section 4.

\section{Observations and Methods} \label{sec:obser methods}
\subsection{Observations}
Our data set is primarily from the New Vacuum Solar Telescope (NVST)\footnote{\url{http://fso.ynao.ac.cn}} (NVST:\citealp{liu14,yan20}) and Atmospheric Imaging Assembly (AIA; \citealp{lem12}) on board the Solar Dynamic Observatory\footnote{\url{https://sdo.gsfc.nasa.gov}} (SDO: \citealp{pes12}). The NVST is a vacuum solar telescope with a 985 mm clear aperture located at Fuxian Lake, in Yunnan Province, China. It provides high-resolution observations of prominence at the H$\alpha$ line (6562.8 $\rm\AA$). The H$\alpha$ images are recorded by a tunable Lyot filter with a bandwidth of 0.25 $\rm\AA$. The H$\alpha$ blue wing (-0.4 $\rm\AA$), H$\alpha$ center, and H$\alpha$ red wing (+0.4 $\rm\AA$) images are used in this study. The field of view of H$\alpha$ images is 150$\arcsec$ × 150$\arcsec$, with a 45s cadence and a spatial resolution of 0.165$\arcsec$ per pixel. These data are calibrated from Level 0 to Level 1, and are dark-current subtracted and flat-field   corrected, before a speckle masking method is used to reconstruct the calibrated images from Level 1 to Level 1+ \citep{xia16}. All NVST H$\alpha$ observations are normalized by the quiet Sun in line with each other based on a cross-correlation algorithm \citep{yang15b}. The AIA instrument on SDO can provide seven simultaneous full-disk EUV images with a pixel scale of 0.6$\arcsec$ and a cadence of 12 seconds. The SDO/AIA and NVST observations are carefully co-aligned by matching specific features observed simultaneously in both the AIA 171 $\rm\AA$ and NVST H$\alpha$ center channels. We mainly use the 211, 193, and 171 $\rm\AA$ channels to study the evolution and dynamic characteristics of bubbles and plumes and 94, 131, 171, 193, 211, 335 $\rm\AA$ images for the DEM calculation. In addition, EUV images from the Extreme Ultraviolet Imager (EUVI: \citealp{wue04}) on board the Solar Terrestrial Relations Observatory-Ahead\footnote{\url{https://stereo.gsfc.nasa.gov}} (STEREO-A: \citealp{kai08}) are used in this study. We mainly employ 304 $\rm\AA$ and 195 $\rm\AA$ images with spatial resolutions of 1.59$\arcsec$ per pixel and cadences of 10 and 2.5 minutes, respectively.
\subsection{Methods}\label{methods}
\subsubsection{Differential emission measures method}
To diagnose the thermodynamical natures of bubble and plumes, we employ the DEM method to derive the emission measure (EM) distribution depending on temperature (T) \citep{su18}. The DEM method is a modified version of the sparse inversion technique for thermal diagnostics developed by \cite{che15}. The DEM solutions, which are derived from several popular DEM codes (such as the regularized inversion code \citep{han12}, the XRT$\_$dem$\_$iterative2 \citep{web04}, and the sparse inversion code \citep{che15}), show significant emissions at high temperatures, which is not consistent with observations.  The modified method provides some useful settings of basis functions and tolerances with which to better constrain the plasma DEMs at high temperatures using AIA data only. The maps of six EUV wavelengths including 94, 131, 171, 193, 211, and 335 $\rm\AA$ obtained from SDO/AIA are used in this method. In order to improve the signal-to-noise ratio (S/N), we average intensities of images within 1 minute for the DEM method. The uncertainties of EM solutions are estimated by 200 Monte Carlo simulations with the uncertainties on AIA intensity obtained using the ``aia$\_$bp$\_$estimate$\_$error.pro'' routine in SolarSoftWare (SSW:\citealp{fre98}). The total EM is defined as 
\begin{equation}
{\rm EM }=\sum {\rm EM}_T=\int n_e^2dl,
\end{equation}
where $n_e$ and $l$ are the electron number density and the optical depth along the line of sight (LOS), respectively. The EM-weighted average temperatures ($T_{em}$) are also estimated by following the formula \citep{su18}:
\begin{equation}
T_{em}=\frac{\sum{({\rm EM}_T\cdot T})}{\sum {\rm EM}_T}.
\end{equation}

\subsubsection{Doppler signals}
Based on observations of H$\alpha$ blue (-0.4 $\rm\AA$) and red (+0.4 $\rm\AA$) wings, the Doppler signals (D$_s$) in the prominence can be calculated using the following formula:
\begin{equation}
{\rm D}_s=\frac{I_{red}-I_{blue}}{I_{red}+I_{blue}},
\end{equation}
where $I_{blue}$ and $I_{red}$ are the emission intensities of the blue wing (H$\alpha$ -0.4 $\AA$) and the red wing (H$\alpha$ +0.4 $\AA$) of the H$\alpha$ line, respectively. The Doppler signal is a proxy of the LOS velocity field in chromospheric temperature \citep{tsi00}. In the formulation, positive values of D$_s$ denote the plasma away from the observer (redshift) while negative values of D$_s$ denote the plasma close to the observer (blueshift).
\section{Results}
\subsection{Overview of the bubble of interest}
In 2021 April 14, a prominence appeared at the northeast limb of the solar disk as viewed  from Earth, corresponding to the bottom of a coronal cavity or the lower side of large-scale coronal loops (see Fig.1 (a)). Figure 1 (a) shows an overview of the prominence in SDO/AIA 211 $\AA$ at 06:08 UT. Viewed from STEREO-A, this prominence was observed from the southwest and became an on-disk quiescent filament (see Fig.1 (b) and (c)). The angle between SDO and STEREO-A was 53.6$\degr$ at this moment (see the blue box in the Fig.1 (c)). Figure 1 (d) displays the fine structures of the prominence in the H$\alpha$ line center observed by NVST. In order to better reveal the structure of the prominence, we rotate the map and reverse the intensity of H$\alpha$ observations (see Fig.1 (d)). The prominence consists of three parts, including two helmet-like structures and a sparse arch-like structure. The bubble of interest resides at the bottom of the middle helmet-like structure that mainly consists of vertical thin threads, and is marked by the white arrow. The bubble has a sharp arch-like boundary below the prominence with the projected area of about 35 ${\rm Mm}^2$ at 06:08 UT. Panels (b)-(c) show observations of a prominence from a different viewing angle, at 304 $\AA$ and 195 $\AA$ from STEREO-A/EUVI, respectively. Based on the world coordinate system (WCS: \citealp{gre02,tho10}) and comparing distinct characteristics between SDO and STEREO-A, we identify the bubble marked by the white circle in the STEREO-A/EUVI observation. We find that the bubble exhibits a brightness of cusp shape in 195 $\AA$ and is faded in comparison at 304 $\AA$ because of strong absorption by the prominence in the He II line. Combined with the observations from two different viewing angles, we can identify that the bubble is beneath the main body of the prominence. More importantly, based on the brightness in the bubble observed by STEREO-A, we can certify that the bubble is filled with hot plasma (exciting strong emissions in the EUV line) rather than being a void structure. Panels (e) and (f) exhibit the prominence in 171 $\AA$ and 193 $\AA$ observations from SDO/AIA. The prominence appears as a dark object because of the absorption of the prominence plasma at  171 $\AA$ and 193 $\AA$. However, the bubble appears as a brightening structure in these two channel observations.

\begin{figure*}
\centering
\includegraphics{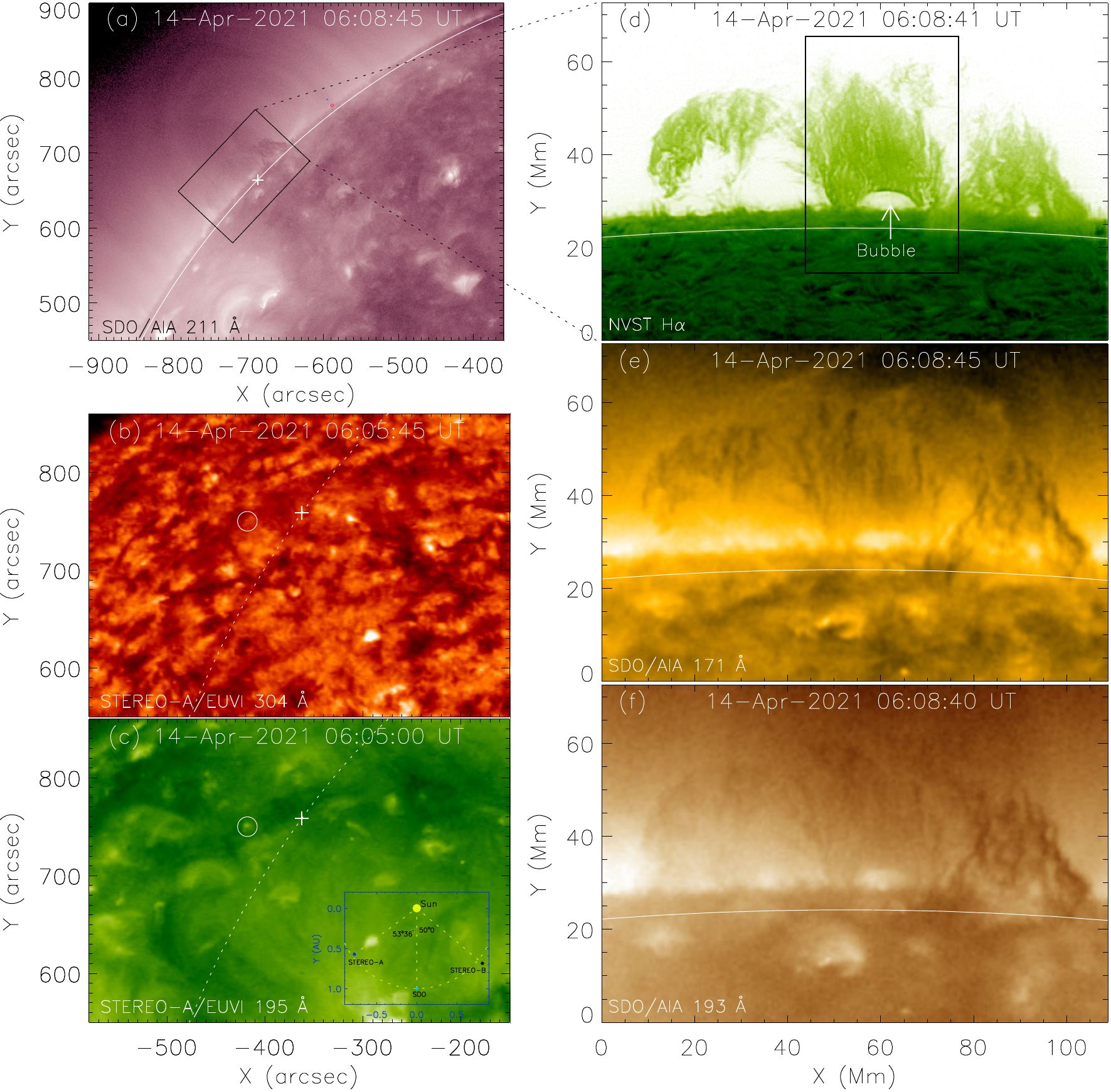}
\caption{Overview of the prominence. (a) SDO/AIA 211 $\AA$ observation. Solar limb is plotted using a white line and the black box outlines the field of view of panels (d)-(f). (b)-(c) Observations of STEREO-A/EUVI in 304 $\AA$ and 195 $\AA$. The white dotted line indicates the solar limb observed from SDO and the white circle marks the position of the bubble. The blue box in the bottom-right corner shows the relative locations of STEREO-A, STEREO-B, and SDO at 06:05 UT. The coordinates of the map in panels (b)-(c) are the heliographic coordinates from STERO-A. The white plus symbols in panels (a), (b), and (c) mark the same point on the surface from different viewpoints, which is derived by the solarsoft WCS Routines in SSW. (d) H$\alpha$ observation from NVST. The black box outlines the field of view of Figs. 2 and 3. The white arrow points to the bubble of interest. (e) SDO/AIA 171 $\AA$ observation. (f) SDO/AIA 193 $\AA$ observation. The solar limb is outline by white lines in panels (d)-(f).}\label{fig1}
\end{figure*}

\subsection{Formation processes of two large plumes}
\begin{figure*}
\centering
\includegraphics{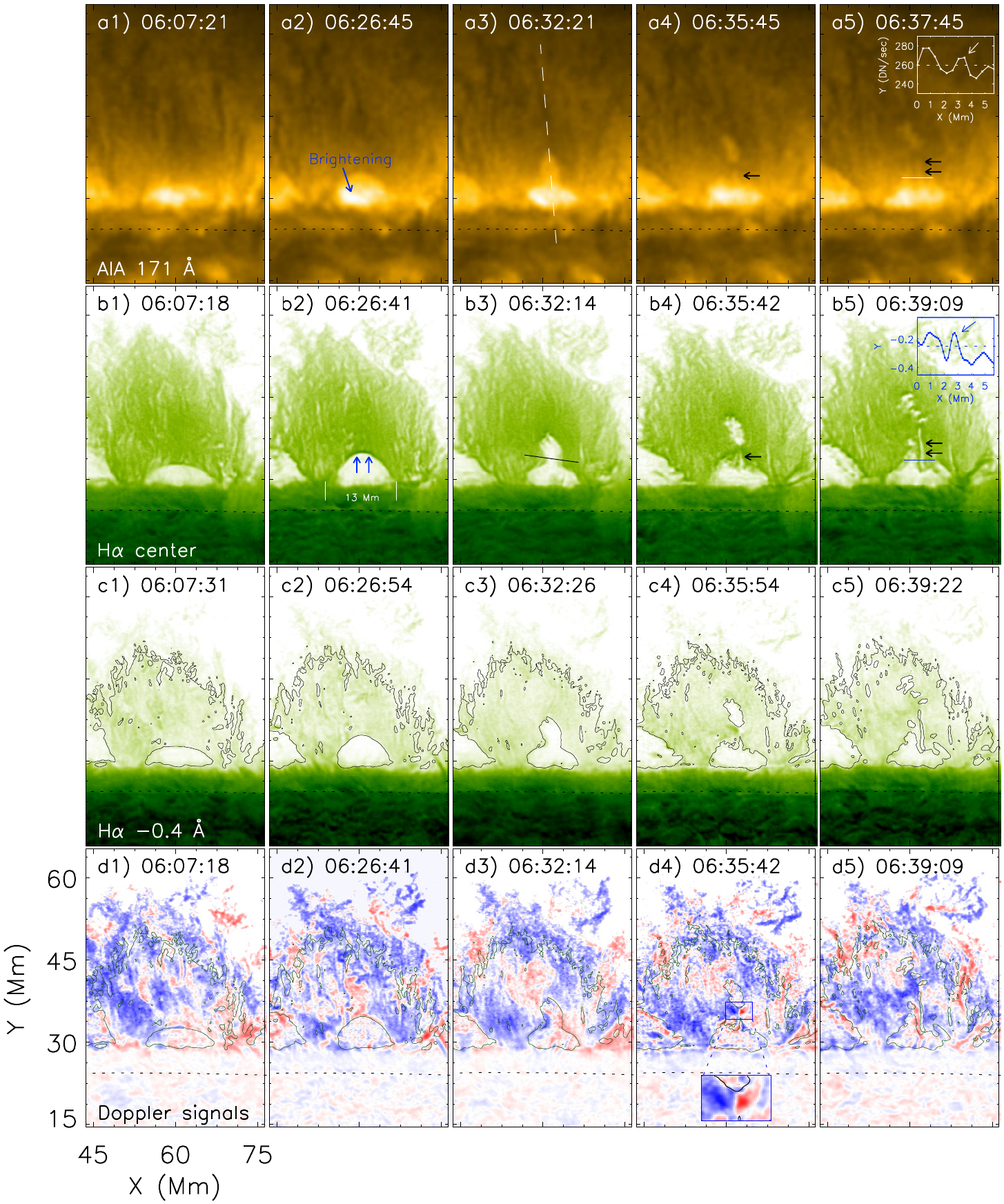}
\caption{Evolution of the first large plume. (a1) - (a5) SDO/AIA 171 $\AA$ observations. The blue arrow in panel (a2) points to the brightening in the bubble. White dashed line in panel (a3) marks the position of time--distance diagrams in Fig.4 (a) - (d). The white curve in panel (a5) depicts the variation of normalized intensity of SDO/AIA 171 $\rm\AA$ along the white line in panel (a5). (b1) - (b5) NVST H$\alpha$ center observations. The black line in panel (b3) marks the slit position of time--distance diagram in Fig.4 (e). The blue curve in panel (b5) depicts the variation of normalized intensity of the inverse H$\alpha$ center along the blue line in panel (b5). (c1) - (c5) NVST H$\alpha$ blue wing (-0.4 $\AA$) observations. (d1) - (d5) Corresponding Doppler signals. The contours in panels (c1) - (c5) and (d1) - (c5) outline the corresponding inverse H$\alpha$ center with a level of 0.25 of the normal intensity. The black horizontal arrows mark the slender void thread in panels (a4)-(a5) and (b4)-(b5). }\label{fig2}
\end{figure*}

Figure \ref{fig2} shows the formation processes of the first large plume originating from the bubble during the period from 06:07 UT to 06:39 UT. Panels (a1) - (a5), (b1) - (b5), and (c1) - (c5) display the formation process of the first large plume in SDO/AIA 171 $\AA$, the NVST H$\alpha$ center, and  the blue wing (H$\alpha$ -0.4 $\AA$), respectively. The corresponding Doppler signals are exhibited in panels (d1) - (d5). An sharp arch-like boundary between the bubble and overlying prominence can be identified at around 06:07 UT. The bubble corresponds to a void region in the H$\alpha$ line (see panels (b1) and (c1)) and a brightening structure at 171 $\AA$ (see panel (a1)). At the early phase, the bubble expands gradually and rises away from the surface (see the first and second columns of Fig.\ref{fig2}). During this period, there are some brightenings occurring in the bubble in 171 $\rm \AA$ observations (see panel (a2)). Around 06:26 UT, the bubble grows into a quasi-semicircle with a width of about 13 Mm. Simultaneously, some enhanced emission of H$\alpha$ center appear at the high part of the bubble--prominence interface, which is marked by blue arrows in panel (b2). This striking feature was also found in previous studies \citep{ber10,awa19,xue21}, which could be interpreted as accumulated chromospheric materials in that region. We find that these accumulated chromospheric plasma are dominated by redshifted signals (see panel (d2)). After that, the bubble becomes unstable and changes shape. At around 06:32 UT, a large flame-like plume runs out from the top of the bubble. Meanwhile, two side edges connecting the bubble and plume approach each other while the left and right side edges are dominated by blue- and redshifts, respectively (see the breakup region in panel (d4)). At 06:35 UT, the large plume separates from the bubble and continues to ascend while some prominence material falls down into the bubble (see panel (b4)). The plume is brighter than the background prominence at 171 $\AA$, which is comparable with the boundary between bubble and prominence in brightness (see panels (a3) and (a4)). As the plume ascends, some material accumulates in the front of the plume, and exhibits a turbulent shape (see panels (b3) - (b5)). Interestingly, a slender void thread connecting the bubble and ascending plume is seen in the H$\alpha$ observation (marked by black arrows in panels (b4) and (b5)), which corresponds to the bright structure at 171 $\AA$  (see panels (a4) and (a5)). The left and right sides of this void thread are dominated by strong blue- and redshifted Doppler signals, respectively (see the zoomed-in region in panel (d4)). At around 06:39 UT, the plume continues to rise and evolves into several fragments. Simultaneously, the slender void thread elongates up to 8 Mm in length in H$\alpha$. Its width is about 0.60$\pm$0.12 Mm (five pixels) in H$\alpha$ (see the blue line in panel (b5)) and 0.87$\pm$0.44 Mm (two pixels) in the 171 $\rm\AA$ observations (see the blue line in panel (a5)). The uncertainties come from the spatial resolutions of the observing instruments. This slender void thread exists for about 7 minutes. After that, the plume and slender thread disperse into the prominence.

\begin{figure*}
\centering
\includegraphics{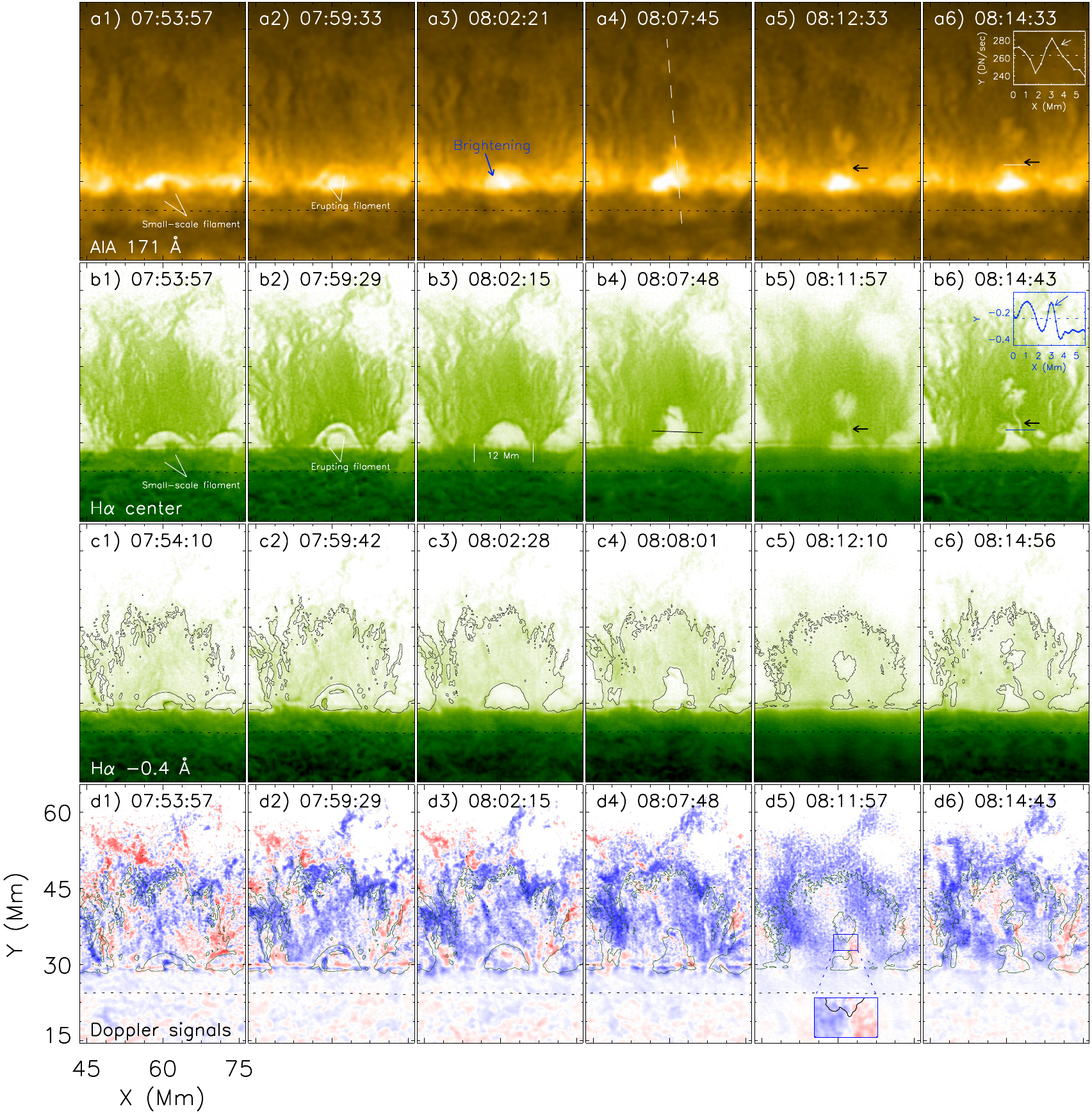}
\caption{Evolution of the second large plume. (a1) - (a6) SDO/AIA 171 $\AA$ observations. The white dashed line in panel (a4) marks the position of the time--distance diagrams in Fig.4 (a) - (d). The white curve in panel (a6) depicts the variation in normalized intensity of SDO/AIA 171 $\rm\AA$ along the white line in panel (a6). (b1) - (b6) NVST H$\alpha$ center observations. The black line in panel (b4) marks the slit position of the time--distance diagram in Fig.4 (f). The blue curve in panel (b6) depicts the variation of normalized intensity of inverse H$\alpha$ center along the blue line in panel (b6). (c1) - (c6) NVST H$\alpha$ blue wing (-0.4 $\AA$) observations. (d1) - (d6) Corresponding Doppler signals. The contours in panels (c1) - (c6) and (d1) - (c6) outline the corresponding inverse H$\alpha$ center with a level of 0.25 of the normal intensity. The black horizontal arrows mark the slender void thread in panels (a5)-(a6) and (b5)-(b6). }\label{fig3}
\end{figure*}

Figure \ref{fig3} displays the evolution of the second large plume originating from the same bubble as the first plume. In the initial stage, the bubble is small and an active small-scale filament can be identified at the bottom of the bubble at around 07:53 UT (see Fig.\ref{fig3} (a1) and (b1)). As this active small-scale filament erupts, the bubble expands and gets higher (see the second column of Fig.\ref{fig3}). The active small-scale filament exhibits strong blueshifted signal during its eruption (see Fig.\ref{fig3} (d1) and (d2)). As in the first large plume, some brightenings are also found in the bubble in the 171 $\rm \AA$ observations during the eruption of the small-scale filament (see panel (a3)). At around 08:02 UT, the erupting small-scale filament merges into the boundary between bubble and prominence while the bubble also grows into a quasi-semicircle with a width of about 12 Mm. In the course of bubble expansion, some enhanced emission of H$\alpha$ center can also be found at the high part of bubble--prominence interface. After that, the bubble experiences a similar episode to the previous one. The bubble becomes unstable and begins to collapse. At 08:07 UT, a large plume comes out from the top of the bubble while the two side edges of the plume approach each other. At around 08:12 UT, the plume separates from the bubble and keeps rising. Simultaneously, we also identify a slender void thread marked by black arrows in panels (b5) and (b6), connecting the bubble and ascending plume in the H$\alpha$ observation despite the poor seeing  (see panel (b5)). The slender void thread is also corresponding to the bright structure in the 171 $\AA$ observation (marked by the corresponding black arrows in panel (a5) and (a6)). Left and right sides of this void thread are also dominated by strong blueshifted and redshifted Doppler signals, respectively (see the zoomed-in region in panel (d5)). At around 08:14 UT, the plume continues ascending and evolves into several fragments as in the first plume. The slender thread is also elongated and distorted. Additionally, its width was measured to be about 0.60$\pm$0.12 Mm (five pixels) in H$\alpha$ (see the blue line in panel (b6)) and 1.31$\pm$0.44 Mm (three pixels) in 171 $\rm\AA$ observations (see the blue line in panel (a6)). After that, the plume lifts and disperses into the prominence while the slender void thread breaks up into the prominence. This slender thread also lives for about 7 minutes. At around 08:27 UT, the bubble breaks up completely. It is worth emphasizing that there were many small irregular plumes in addition to these two large plumes, initially occurring at the bubble--prominence interface during the formation of these two large plumes. The appearance of these small plumes is similar to that of some previously reported plumes \citep{ber08,ber10,ber11,xue21}. These small plumes were much smaller  in size than the two large plumes, and occurred in random places at the bubble--prominence interface instead of the bubble top. 

\begin{figure*}
\centering
\includegraphics{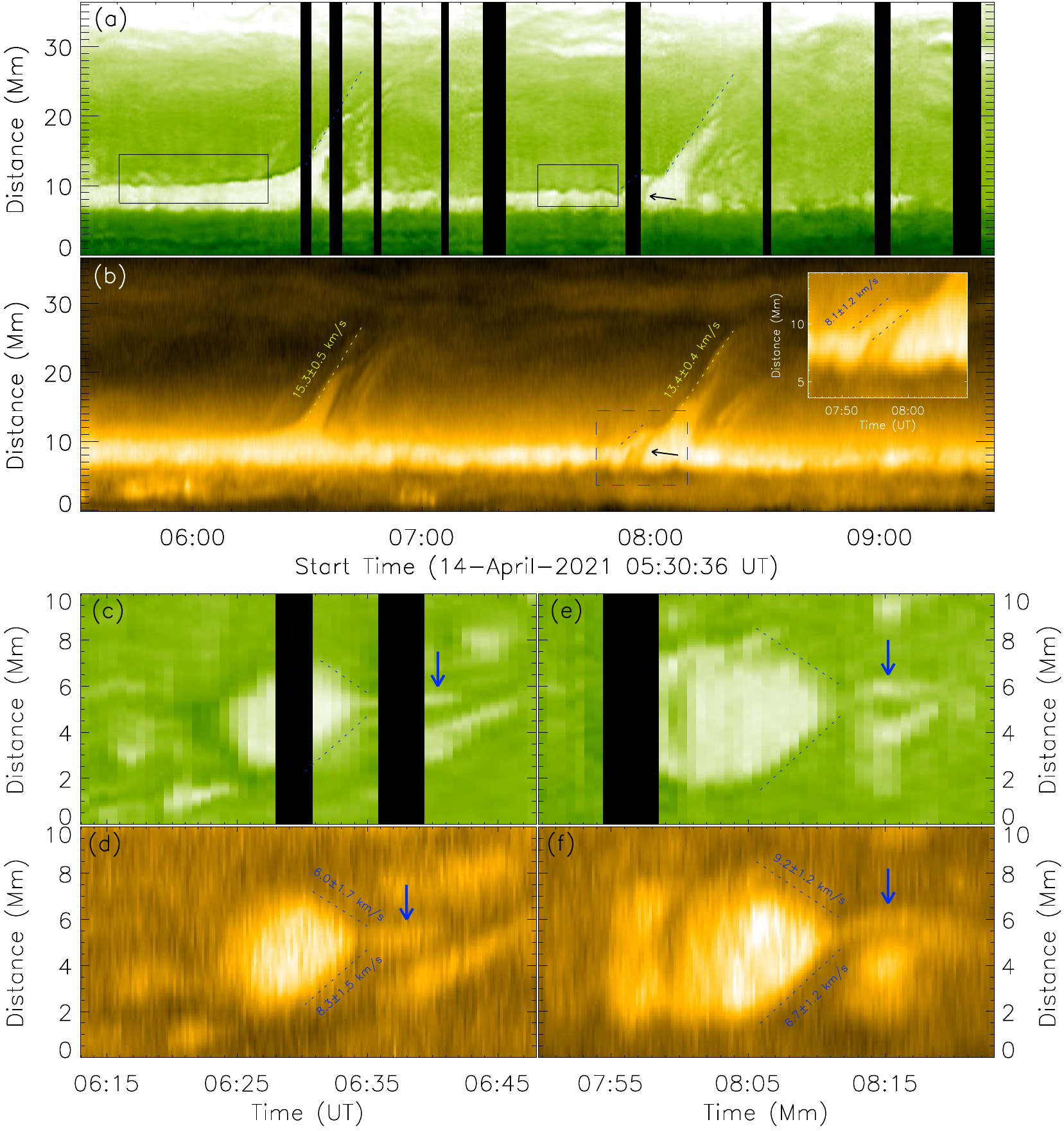}
\caption{Time--distance diagrams derived by a sequence of images. (a) NVST time--distance diagram along white dashed lines in Fig.2 (a3) and Fig.3 (a4). (b) Corresponding SDO/AIA 171 $\AA$ time--distance diagram. The inset time--distance diagram in the top right is a zoomed-in image of the region marked by the blue dashed box. (c) - (d) Time--distance diagram derived by NVST H$\alpha$ and SDO/AIA 171 $\AA$ images along the black line in Fig.\ref{fig2} (b3). (e) - (f) Time--distance diagram derived by NVST H$\alpha$ and SDO/AIA 171 $\AA$ images along the black line in Fig.\ref{fig3} (b4). The uncertainties on the velocities are estimated as the error in the length measurement which is considered to be about 0.6 arcsec (about one pixel length of SDO/AIA).}\label{fig4}
\end{figure*}

\begin{figure*}
\centering
\includegraphics{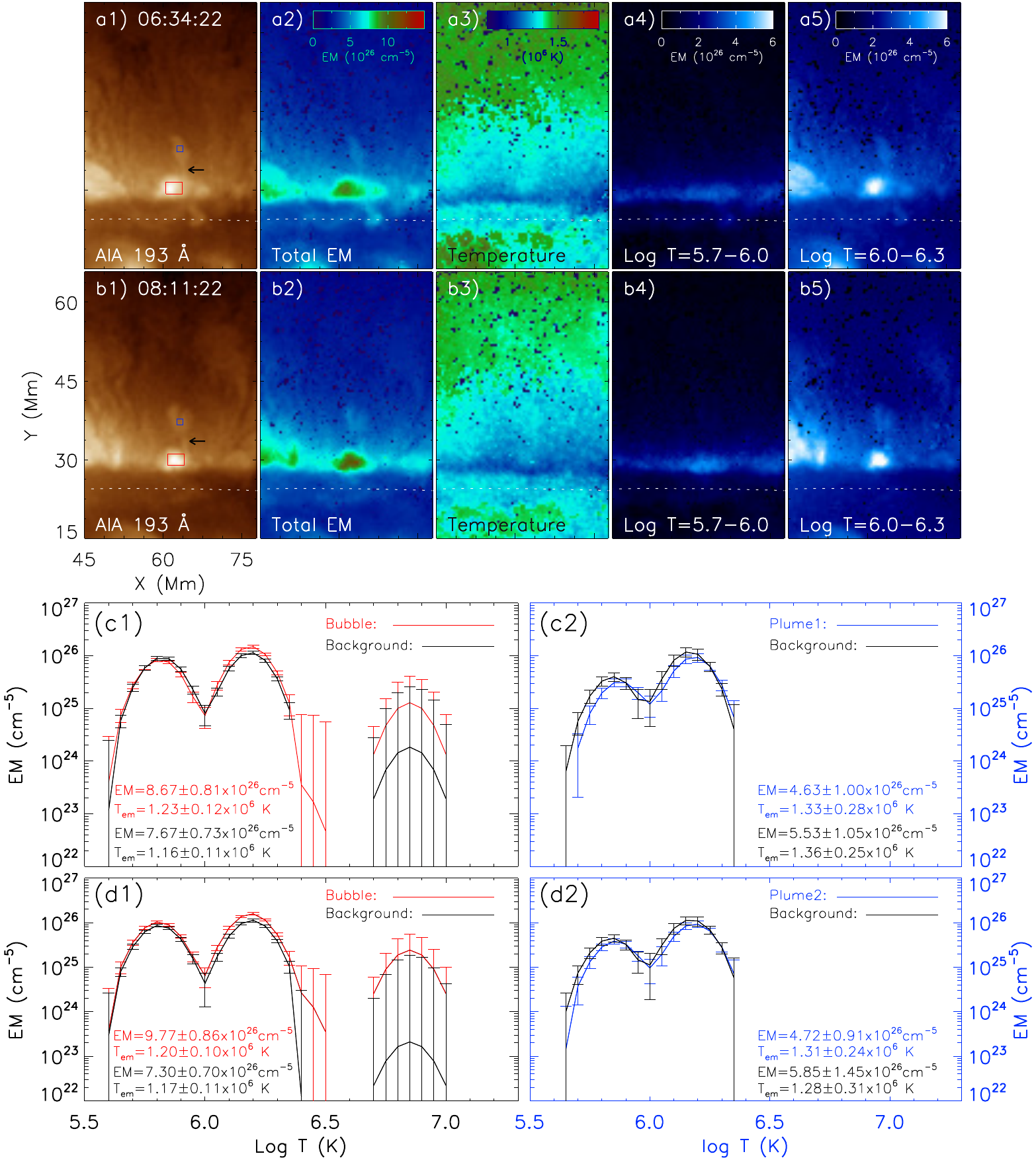}
\caption{Thermal characteristics in two large plumes. (a1) - (a5) 193 $\AA$ image, total emission measures, EM average temperature, emission measures at log T = 6.0 - 6.3, and emission measures at log T = 6.0 - 6.3 at 06:34:22 UT. (b1) - (b5) Same as panels (a1) - (a5) at 08:11:22 UT. (c1) - (c2) Average EM curves in the red and blue boxes of panel (a1) for the first large plume. (d1) - (d2) Average EM curves in the red and blue boxes of panel (b1) for the second large plume. The corresponding background EM curves are derived from the corresponding quiet Sun away from the prominence (red and blue boxes in Fig.1 (a)). The uncertainties were estimated using the Monte-Carlo experiment.}\label{fig5}
\end{figure*}
\subsection{Dynamic properties}
In order to investigate the kinetic characteristics of these two large plumes, we made two time--distance maps derived using time-series images of NVST H$\alpha$ center and SDO/AIA 171 $\AA$, which are along the same path outlined by the white dashed lines in Figs. 2 (a3) and 3 (a4). Figure \ref{fig4} (a) - (b) exhibit the time-distance maps in H$\alpha$ center and 171 $\AA$, respectively. Similar behaviors of bubble and plume are seen in these two time--distance maps. Before the occurrence of the first large plume, the bubble--prominence interface ascends gradually. This indicates that the bubble is ascending before the first large plume. Before the second large plume, the bubble--prominence interface also lifts up sharply, accompanied by some dark material (marked by the black arrows in panels (a) and (b)) related to the small-scale filament eruption. Before the bubble--prominence interface rises up to about 12 Mm, the bubble--prominence interface and the front of the rising small-scale filament lifts up at a constant speed of about 8.1$\pm$1.2 km/s (see the inset time--distance diagram panel (b)). The two large plumes travel almost at a constant speed. The average ascending speeds of the two large plumes are about 15.3$\pm$0.5 km/s and 13.4$\pm$0.4 km/s, respectively. In addition, before their appearance, the bubble--prominence interface experiences several disturbances marked by two black boxes in panel (a). We also used the time--distance maps to derive the approaching speeds of the two-side edges connecting the bubble and plume during the formation of the plumes. Figure \ref{fig4} (c) - (d) shows the H$\alpha$ and AIA 171 $\AA$ time--distance maps for the first large plume along the black solid line in panel (b3) of Fig.2. The approaching speeds of the left and right sides are about 8.3$\pm$1.5 km/s and 6.0$\pm$1.7 km/s, respectively. Thus, the relative approaching speed is about 14.3$\pm$3.2 km/s. Figure \ref{fig4} (e) - (f) shows the H$\alpha$ and AIA 171 $\AA$ time--distance maps for the second large plume along the black solid line in panel (b4) of Fig.\ref{fig3}. The approaching speeds of the left and right sides are about 6.7$\pm$1.2 km/s and 9.2$\pm$1.2 km/s, respectively. Thus, their relative approaching speed was about 15.9$\pm$2.4 km/s. Therefore, both large plumes have almost the same approaching speeds as the two side edges during their formation. Another feature can be identified in these four time--distance diagrams, namely the cross-section of the slender void thread. For both large plumes, the signals of slender void threads, marked by blue arrows in panels (c)-(f), can be distinguished in both H$\alpha$ and AIA 171 $\AA$ time--distance diagrams.

\subsection{Thermal properties}
Figure \ref{fig5} shows the thermal characteristics derived using the DEM method in these two large plumes. Panels (a1) - (a5) display the 193 $\AA$ image, distribution of total emission measures, EM-weighted average temperature, and emission measures at log T = 5.7 - 6.0 and log T = 6.0 - 6.3 for the first large plume at 06:34:22 UT, respectively. Panels (b1) - (b5) are the same images for the second large plume at 08:11:22 UT. The bubble, plume, and slender void thread can be seen at that time. The slender void threads are marked by the black horizontal arrows in panels (a1) and (b1). The bubble exhibited strong emission measures (see panels (a2) and (b2)). The distribution of the EM-weighted average temperature, which is on the order
of 10$^6$ K, is complex. The bubble, plume, and slender void thread have strong emission measures at log T=6.0 - 6.3 (see panels (a5) and (b5)), while the bubble also has strong emission measures at log T=5.7 - 6.0 (see panels (a4) and (b4)). This implies that these structures in H$\alpha$ have strong emission measures at log T = 5.7 - 6.3. Figure \ref{fig5} (c1) and (c2) and (d1) and (d2) show the mean emission measure profiles calculated using the DEM solution within different regions representing the bubble and plumes for two large plumes, respectively. Additionally, we estimated the corresponding emission measure profile in the neighboring quiet corona away from the prominence marked by the red and blue boxes in Fig.\ref{fig1} (a). The emission measures in these structures are also found to be mainly dominated by the temperature at log T = 5.7-6.0. They are the subject of large uncertainties at log T =6.4-7.0, and therefore we only consider the EMs with uncertainties smaller than half of their value. For the first large plume at around 06:34 UT, the total EM and $T_{em}$ were estimated to be 8.67$\pm$0.81 $\times$ 10$^{26}$ cm$^{-5}$ and 1.23$\pm$0.12 $\times$ 10$^6$ K in the bubble, respectively (see panel (c1)). Meanwhile, the EM and $T_{em}$ in the background were 7.67 $\pm$0.73 $\times$ 10$^{26}$ cm$^{-5}$ and 1.16$\pm$0.11 $\times$ 10$^6$ K, respectively. On the other hand, the EM and $T_{em}$ in the plume were estimated to be,  respectively, 4.63$\pm$1.00 $\times$ 10$^{26}$ cm$^{-5}$ and 1.33$\pm$0.28 $\times$ 10$^6$ K, while the corresponding EM and $T_{em}$ in the background were 5.53$\pm$1.05 $\times$ 10$^{26}$ cm$^{-5}$ and 1.36$\pm$0.25 $\times$ 10$^6$ K (see panel (c2)). For the second large plume at around 08:11 UT, the EM and $T_{em}$ in the bubble were estimated to be,
respectively, 9.77$\pm$0.86 $\times$ 10$^{26}$ cm$^{-5}$ and 1.20$\pm$0.10 $\times$ 10$^6$ K, while the corresponding EM and $T_{em}$ in the background were 7.30$\pm$0.70 $\times$ 10$^{26}$ cm$^{-5}$ and 1.17$\pm$0.11 $\times$ 10$^6$ K (see panel (d1)). Panel (d2) shows the profile of EM in the second large plume. In the second large plume, the EM and $T_{em}$ were estimated to be,
respectively, 4.72$\pm$0.91. $\times$ 10$^{26}$ cm$^{-5}$ and 1.31$\pm$0.24 $\times$ 10$^6$ K, while the corresponding EM and $T_{em}$ in the background were 5.85$\pm$1.45 $\times$ 10$^{26}$ cm$^{-5}$ and 1.28$\pm$0.31 $\times$ 10$^6$ K. In general, we find that the EM in the bubble is in excess of the background, and the EM-weighted average temperature in the bubble of $\sim$ 1.2 $\times$ 10$^6$ K is comparable to that of the background. However, the EM in the plumes is lower than that of the background while the $T_{em}$ in the plumes of $\sim$ 1.3 $\times$ 10$^6$ K is also comparable to that of the background. This suggests that these large plumes have a temperature  comparable to that of the bubble but show much lower EMs than the lower surface of the bubble. This shows that the bubble and plume consist of hot plasma at about 1 - 2 $\times$ 10$^6$ K, which is two orders of magnitude hotter than the overlying prominence \citep{par05}. Assuming that the depth ($l$) of the emitting plasma is equal to the   widths of the boxes, the electron number density ($n_e$) can be calculated using the formula: $n_e=\sqrt{{\rm EM}/l}$. Based on the above assumption and derived parameters, the average densities of bubble and plume can be estimated to be about 1.7 $\times$ 10$^9$ cm$^{-3}$ and 2.0 $\times$ 10$^9$ cm$^{-3}$, respectively.
\begin{figure*}
\centering
\includegraphics{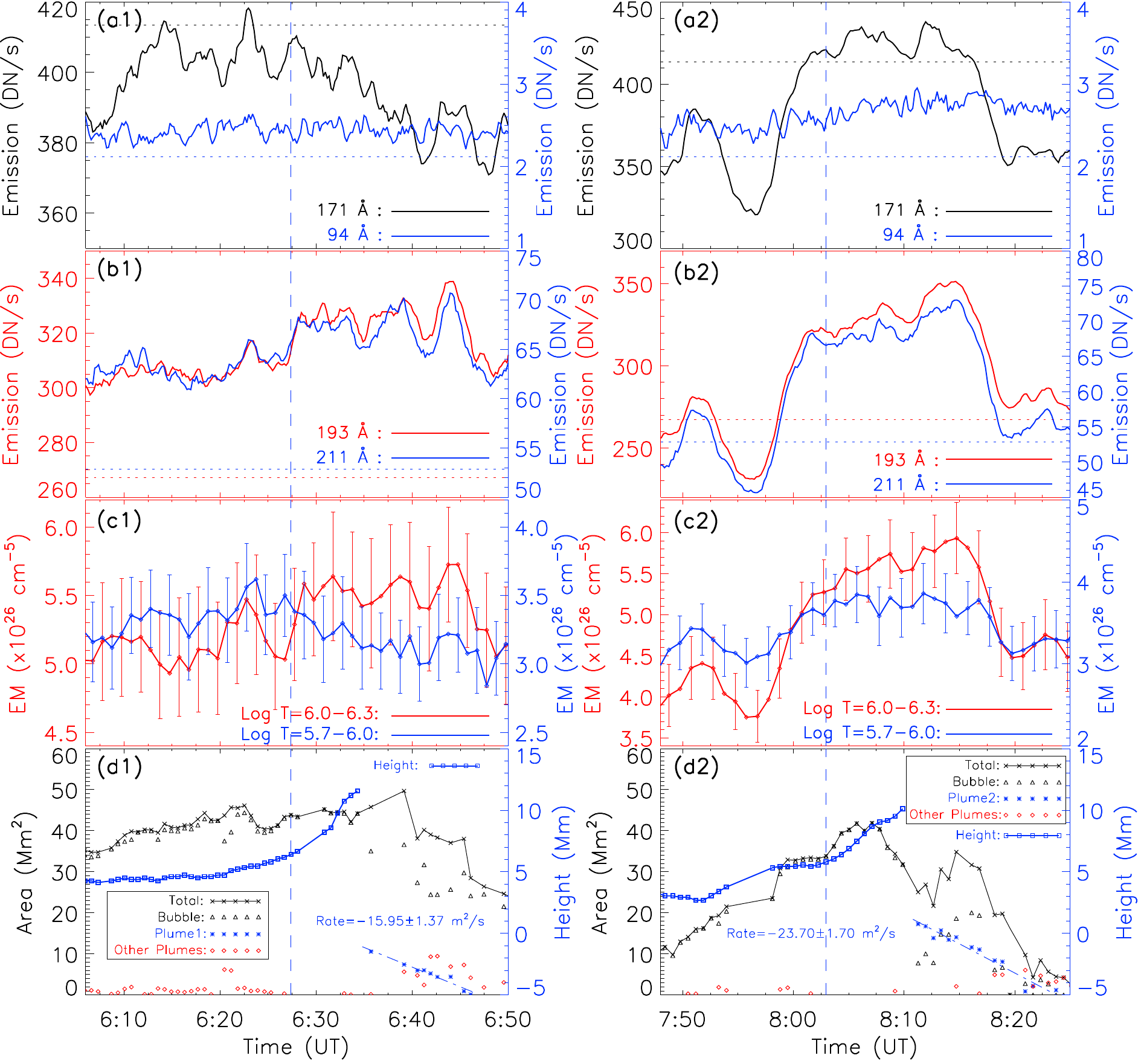}
\caption{Evolution of different parameters in the bubble and plumes. (a1-b1, c1, d1) Variations of different SDO/AIA normalized intensities by exposure duration, EM in the red box of Fig.5 (a1), projected areas, and the height for the first large plume. (a2-b2, c2, d2) Variations of different SDO/AIA normalized intensities by exposure duration, EM in the red box of Fig.5 (b1), projected areas, and the height for the second large plume. The blue vertical dashed line denotes the onset of instability in the bubble top for each large plume. The horizontal dotted lines in panels (a1)-(b1) and (a2)-(b2) denote the corresponding average normalized intensity of background at different wavelengths. The uncertainties in panels (c1)-(c2) are derived directly from the errors of Fig.5 (c1)-(d1) estimated by the Monte-Carlo experiment.}\label{fig6}
\end{figure*}

\subsection{Variations of different physical parameters}
In order to investigate the triggering mechanisms of these two large plumes, we calculated different physical parameters in the bubble during the formation of these two large plumes. In addition, we used H$\alpha$ center images to derive the aspects of the bubble and plume, where we use the contours of normal intensity at the level of 0.25 plotted in Figs. \ref{fig2} and \ref{fig3} as the boundaries of bubble and plumes. The dashed vertical line denotes the onset of instability at the top of the bubble for each large plume (around 06:27 UT for the first plume and around 08:03 UT for the second plume, see the animations 1 and 2). Figure \ref{fig6} shows time variations of these parameters related to two large plumes. For the first large plume, at the early phase, the AIA/SDO EUV emissions in the bubble display a slow enhancement trend at 171, 193, and 211 $\AA$. It is worth noting that the radiation emissions at 171 $\rm \AA$ in the bubble are slightly smaller than the background but at 193 and 211 $\rm \AA$ these strongly exceed those of the background. The radiation emissions in 94 $\rm\AA$ remain invariable. Meanwhile, EMs at log T= 5.7- 6.0 and log T =6.0-6.3 also exhibit a slight enhancement. Simultaneously, the projected areas and the height of the bubble also experience a slow increase. It is noted that several small plumes like the ones discovered by SOT on board Hinode also occurred at the bubble--prominence interface during this time. When the projected area and height of the bubble reached about 44 Mm$^2$ and 6.5 Mm at 06:27 UT, the bubble became unstable. We find that the height of the bubble was comparable to half of its width (see Fig. 2 (b2)) at that moment. After that, the height of the bubble increased rapidly, meaning that the plume began to come out from the bubble. At around 06:35 UT, the first plume separated from the bubble, with its projected area of about 10 Mm$^2$. The blue asterisks plot the variation of the projected area of the plume. The projected area of the plume experienced a linear-like decay. With a linear fit, the decaying rate of the projected area is about -15.95 $\pm$ 1.37 m$^2$/s. For the second large plume, unlike the first plume, the EUV emission at the different wavelengths other than 94 $\rm\AA$ increased rapidly during the early phase. This may be related to the eruption of the small-scale filament occurring in the bottom part of the bubble. Both EMs at log T=5.7 - 6.0 and log T=6.0 - 6.3 also enhanced. The EMs at log T=6.0 - 6.3 increased more sharply. Meanwhile, the projected area of the bubble also increased sharply from about 10 Mm$^2$ to 33 Mm$^2$. The height of the bubble also increased from 4 Mm to 6 Mm. At around 08:03 UT, the bubble became unstable and began to generate the second plume. We also find that the height of the bubble was also comparable to half of the width of the bubble (see Fig.\ref{fig2} (b2)) at that moment. At that point, the height of the bubble increased sharply and the plume came out of the bubble. At around 08:11 UT, the second large plume separated from the bubble, with a projected area of about 17 Mm$^2$. As the plume rose up, the plume became smaller and smaller. The projected area of the plume is plotted as the blue asterisks in panel (f). The projected area of this plume also decreased with a linear trend. The linear fit to the projected area decay curve for the second plume in panel (f) shows an initially linear decaying rate of about -23.70 $\pm$ 1.70 m$^2$/s. Altogether, we find enhanced emission of several EUV bands and EMs in the bubble during its growth, which is consistent with the brightenings in Figs. \ref{fig2} (a2) and \ref{fig3} (a3). More importantly, as the height of the bubble becomes comparable to half the width of the bubble, the bubble becomes unstable and generates the large plumes.

\section{Summary and Discussion}\label{sec:conclusion}
In this paper, we investigate the successive formation of two large plumes in a solar prominence on 2021 April 14. Using high-resolution observations from NVST and data observed by SDO and the STEREO satellite, we study the kinetic and thermal characteristics of two large plumes and their formation process. Our main results are as follows:

(1) Two large plumes originate from the same bubble, and are associated with the disruption of the bubble. Before the occurrence of each large plume, the bubble expanded and grew into a quasi-semicircular shape. Simultaneously, the emission intensity of EUV bands (including wavelengths of 171, 193, and 211 $\rm\AA$) and EMs also increased in the bubble. The second large plume was associated with an eruption of the small-scale filament in the bubble. When the height of the bubble reached half of its width, the bubble became unstable and began to generate the large plumes.

(2) A novel phenomenon was discovered during the formation of each plume, namely a slender void thread in H$\alpha$ connecting the bubble and plume. The width of this void thread is about 1 Mm. Furthermore, the left and right sides of this slender void thread are dominated by blue- and redshifted Doppler signals, respectively.
 
(3) The large plumes rose up in the body of the  prominence with a constant ascent speed of about 13-15 km/s. During their development, the projected area of the plumes decreased with a linear trend. The decay rates of these two plumes are about -15.95 $\pm$ 1.37 Mm$^2$/s and -23.70 $\pm$ 1.70 Mm$^2$/s, respectively.

(4) The temperature and electron number density of these two large plumes are estimated to be about 1.3 $\times$ 10$^6$ Kelvin and 2.0 $\times$ 10$^9$ cm$^{-3}$, respectively. Compared to the typical prominence, the large plume is two orders of magnitude hotter and one order thinner.

The nature of prominence bubble ---whether it is a void volume or filled with plasma--- is a controversial problem. We provide extremely powerful evidence supporting the hypothesis that the bubble below the prominence is filled with hot and low-density plasma. Because of a lack of stereoscopic observation of prominence bubbles, some authors considered the prominence bubbles to be open windows that are no obstacle to the background coronal radiation \citep{dud12,gun14}. Here, thanks to two viewing angles, the bubble was seen to exhibit bright structure in the 195 $\AA$ band viewed from the top and a void region in H$\alpha$ line viewed from the side, which shows that the emission seen inside bubbles by SDO/AIA at 171, 193, and 211 $\AA$  originates from the internal plasma rather than from behind or  in front of  the bubble. Furthermore, the bubble has a density of about 1.7 $\times$ 10$^9$ cm$^{-3}$ and a temperature of about 1.2 $\times$ 10$^6$ K, similar to the typical quiet Sun (e.g., \citealp{dos97,lan03,war09, lan10}), which is an order of magnitude thinner and two orders hotter than the prominence \citep{par05,bom86}. This demonstrates that the bubble is a hot, low-density volume beneath the cold and dense prominence instead of an open window.

The two large plumes studied here are quite different from the small-scale dark upflow plumes  discovered with Hinode \citep{ber08,ber10,ber11,ber17}, which are more like the large plumes reported previously \citep{ste73, det08}. The plumes studied here\ are larger in size ($\sim$10-20 Mm$^2$) and have a bubble-related structure. We discovered a novel phenomenon during the formation
of each large plume, that is, a slender void thread with a width of about 1 Mm in H$\alpha$ observations. The plumes show strong emission measures at log T = 6.0 - 6.3, meaning that their temperature might be around 1-2 MK. Interestingly, a similar structure was seen in the numerical simulations \citep{hil12}. In the simulations, a low-density, high-temperature region involved in the bubble, which is consistent with our observation, was considered to yield the plume. The narrow dark region was thought to correspond to the channel of transport of hot plasma from the bubble to the upflow plume. During the formation of the plume in the simulations, magnetic reconnection between the rising and falling magnetic field resulted in strong downflow in addition to this narrow dark channel \citep{hil12c}. In our observations, some downflows 
could also be identified during the formation of the large plumes (see Figs.2 (b4) and 3 (b5) and animations 1 and 2). In addition, the two sides of the plume--bubble connection were found to approach each other during the formation of the large plumes. If the bubble is considered to be an arc-like magnetic field structure \citep{dud12, gun14, guo21}, the magnetic fields on the two sides of the slender void thread should be of opposite direction. We found the left and right sides of the slender void thread to be dominated by blue- and redshifted Doppler signals, respectively. This could be explained as the result of the approaching motion in the plane of bubble arc-like magnetic field that is not parallel to the plane of the solar disk, which is also consistent with the location of the small-scale filament (see Fig.\ref{fig3} (a1) and (b1)). Therefore, we consider these slender void threads to correspond to the channel of bubble-plasma upflow with opposite magnetic field edges (strong electric current).

The formation mechanism of plumes is the key problem in understanding prominence plumes. Previous studies considered the magnetic Rayleigh–Taylor (RT) instability to be responsible for the dark turbulent small-scale upflow plumes \citep{ber08, ber10,ber11,hil12, ber17}. This type of small-scale upflow plume, triggered by small perturbations, is often initiated irregularly over any part of the bubble-prominence interface. The magnetic field transverse to gravity has a strong effect on the RT instability and suppresses the RT instability \citep{ryu10,cha61}. However, these two large plumes were formed at the top of the bubble, and are obviously related to its disruption. We find that the bubble becomes unstable and generates the plumes at the point at which the height of the bubble is comparable to half of its width. The dynamics of the bubble should be subject to the Rayleigh-Plesset equation (see Eq.(\ref{eqA1}) in the Appendix A: \citealp{ray17,ple49}). When the bubble expands over its capacity, the bubble should collapse and release its internal gas. With the modeling we present in Appendix A, the bubble is not able to keep its shape and begins to collapse as the bubble grows and its height exceeds half of its width. This is consistent with our observations. Therefore, we suggest that these large plumes are the result of the disruption of the bubble. On the other hand, it is found that the two side edges of the  top of the bubble approach each other after the bubble has become unstable. The large plume then forms. Furthermore, the left and right side edges are dominated by blue- and redshifts (see the breakup region in Figs.\ref{fig2} (d4) and 3 (d5)), respectively. These observational signals strongly suggest that the escape of the large plumes from the bubble is likely by way of magnetic reconnection \citep{par57,swe58, gon16,xue20,xue21a} at the boundary of the bubble. However, we do not find any strong enhancements in SDO/AIA high-temperature bands (such as 94, 131 $\rm\AA$) around the slender void thread corresponding to the magnetic reconnection site. A possible reason for this is that the magnetic reconnection occurring at the boundary is too weak to enhance the intensities of SDO/AIA high-temperature lines.

Another problem that interests us is the driving force behind the disruption of the bubble. From the Rayleigh-Plesset equation (see Eq.(\ref{eqA1})), the driving parameter for the bubble dynamics is  pressure \citep{fra04}. In our study, we show that the radiation of different EUV wavebands and EMs in the bubble become enhanced before the breakup of the bubble, which urges us to believe that the enhancement of thermal pressure in the bubble is the main driver of the disruption of the bubble. When the pressure in the bubble enhances, the bubbles expand its volume. As the bubble expands enough or the magnetic field cannot maintain its integrity, the bubble becomes unstable and begins to break up. When the bubble breaks up, the top half of the bubble is the first part to become distorted because of the weaker magnetic field there. As the bubble breaks up, some internal gas escapes the bubble and forms the plumes which continue to ascend because of their low density. For the second large plume, the main originating source enhancing thermal pressure in the bubble is related to the eruption of a small-scale filament. As the small-scale filament erupts in the bubble, it releases magnetic energy in the form of thermal energy, momentum, and radiative energy \citep{lin00}. However, for the first large plume, the source of the enhancement of thermal pressure is uncertain. It seems to be a slow process, unlike the second plume. There are several disturbances occurring at the bubble--prominence interface before the appearance of the large plumes. This type of disturbance could result from small--scale energy release under the bubble--prominence interface \citep{she15}. Therefore, we suspect that another likely source of the enhancement of thermal pressure is related to small-scale magnetic reconnections \citep{sam19}.

These two large plumes ascend with an approximately constant speed of about 15 km/s, which is comparable to that of the small-scale plumes \citep{ber08,ber10,ber11}. The ascent speed of plumes may be related to the intrinsic attributes of the prominences instead of the scale of the upflowing plumes. In the course of the upflow of plumes through the main body of the prominence  in a constant motion, ignoring the effect of the magnetic nature of the prominence and its accompanying forces, the plume maintains equilibrium by gravitational forces, a buoyant force, and a viscous resistance force:
\begin{equation}
 {(\rho}_{pro}-\rho_{plu})gV\ -\ \frac{1}{2}cS\rho_{pro}v^2=0,\label{eq4}
\end{equation}
where $\rho_{pro}$ and $\rho_{plu}$ are the densities of the prominence and plume, respectively, $V$ and $S$ are the volume and cross-sectional area of the plume, respectively, and $c$ is the coefficient of viscous resistance in the prominence, related to the dynamic viscosity. As $\rho_{pro}\gg \rho_{plu}$, we obtain that $c\approx 2gV / (Sv^2)$. In the rising plumes, if we assume that the ratio of the volume to cross-sectional area in plumes is a constant, the coefficient of viscous resistance in the prominence is inversely proportional to the square of the ascent speed of plumes. It is worth noting that Eq.\ref{eq4} disregards the magnetic nature of the prominence or assumes that the upflow plasma goes along the vertical magnetic field of the prominence. Therefore, considering the above situations, the comparable derived ascent speed of plumes may reveal the approximate coefficient of viscous resistance in different prominences.
\begin{acknowledgements}
The authors are indebted to the NVST, SDO/AIA and STEREO teams for data support. This work is supported by the National Science Foundation of China (NSFC) under grant numbers 12003064, 11873087, 11803085, Yunnan Science Foundation of China (2019FD085), the National Key R\&D Program of China (2019YFA0405000), the CAS “Light of West China" Program under number Y9XB018001, the Open Research Program of the Key Laboratory of Solar Activity of Chinese Academy of Sciences (grant No. KLSA202014),the Yunnan Talent Science Foundation of China (2018FA001), the Yunnan Science Foundation for Distinguished Young Scholars under No. 202001AV070004, and the Key Research and Development Project of Yunnan Province under number 202003AD150019. Dr. Yang Su and Dr. Mark Cheung are acknowledged for the DEM code used to derive the thermodynamical properties of the plasma from the EUV images.
\end{acknowledgements}

\begin{appendix}
\section{Bubble evolution: Rayleigh-Plesset equation}
To describe the dynamic evolution of the bubble, we must consider the gas pressure, nonlinearity, and dissipative effects. In the semi-spherical shape of the bubble, the bubble radius (R(t)) satisfies the following second-order differential equation known as the Rayleigh-Plesset equation:
\begin{equation}
 \rm \rho \left[R\ddot{R}+\frac{3}{2}\dot{R}^2\right] = [P_v -P_\infty (t)]-\frac{2\sigma}{R}-4\mu\frac{\dot{R}}{R},\label{eqA1}
\end{equation}

in which $\rho$ is the density of prominence plasma, and $\dot{R}$ and $\ddot{R}$ are the first- and second-order derivatives of the bubble radius with respect to time.

\begin{figure*}[h]
\centering
\includegraphics[scale=0.7]{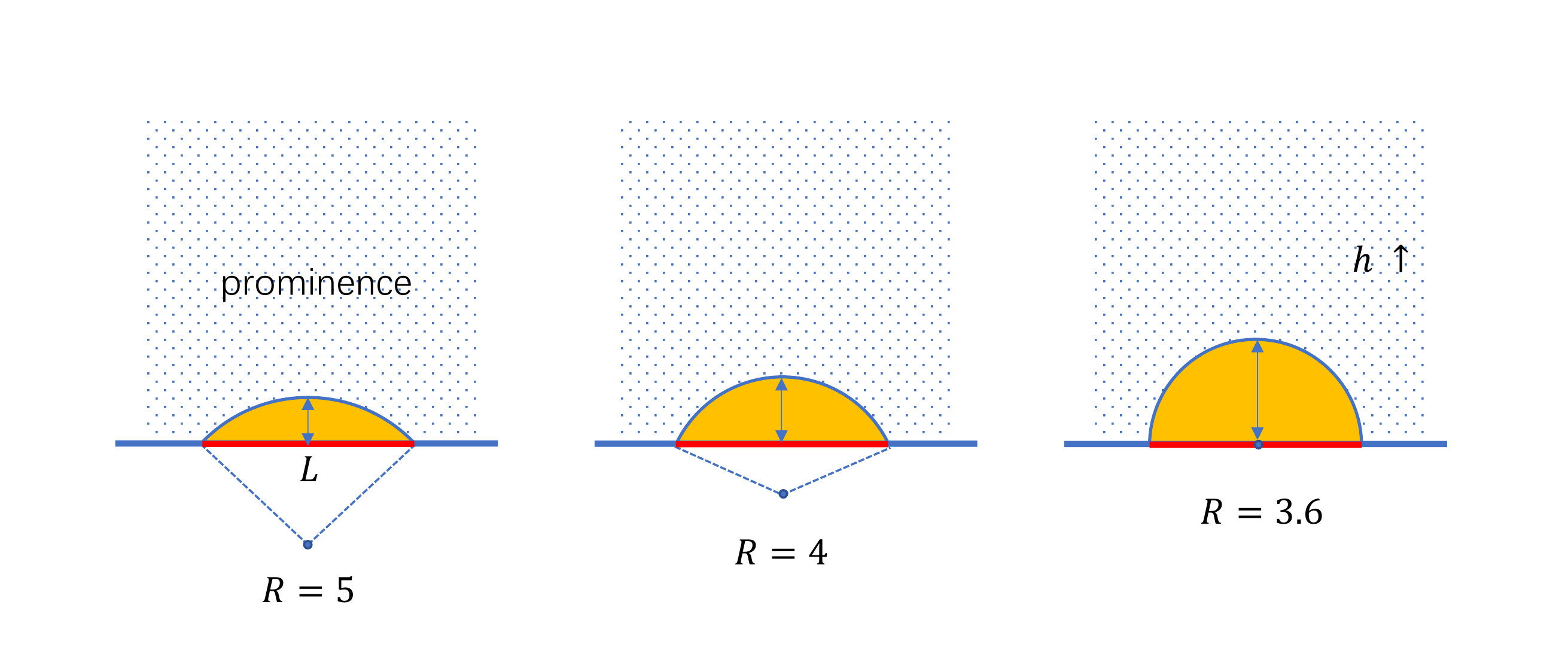}
\caption{Variations of bubble radius and height during the growth of the bubble. The bubble is shown by the orange region.}\label{figA1}
\end{figure*}

On the right-hand side of the equation, the first term ${[P}_v-P_\infty(t)]$, which measures the closeness of the applied pressure to the bubble internal pressure, is the driving term for the bubble evolution. The second term is the surface tension dominated by the magnetic field, $\rm \sigma \sim \frac{B^2}{4\pi}$, and the last term is contributed by effect of dynamic viscosity $\mu$ in the prominence. If we assume that the bubble is a quasi-static evolution, the bubble remains in equilibrium at different heights in the bubble. Then, by setting all time derivatives ($\ddot{R}$, $\dot{R}$) to zero, we get the following equilibrium condition:
\begin{equation}
\rm [P_v -P_\infty (t)]=\frac{2\sigma}{R} \sim \frac{B^2}{2\pi R}.
\end{equation}
If the magnetic field remains invariant during the bubble evolution, the pressure in the bubble should be inversely related to the bubble radius. If a bubble keeps an arcade shape with a constant width of $W$ and a height of $h$, we obtain that:
\begin{equation}
{\rm R}=\frac{1}{2}h+\frac{W^2}{8h}.
\end{equation}
As the bubble expands, the height of the bubble increases while the radius of the bubble decreases (see Fig.\ref{figA1}). According to the above equation, we obtain that:
\begin{equation}
{\rm R}_{min}=\frac{1}{2}W=h_{{\rm R}_{min}},
\end{equation}
which suggests that as the bubble height exceeds half its width, the bubble cannot keep its shape and begins to collapse. This scenario is consistent with our observations.
\end{appendix} 

\begin{thebibliography}{}
\bibitem[Ahn et al.(2010)]{ahn10} Ahn, K., Chae, J., Cao, W., et al.\ 2010, \apj, 721, 74. 
\bibitem[Awasthi \& Liu(2019)]{awa19} Awasthi, A.~K. \& Liu, R.\ 2019, Frontiers in Physics, 7, 218. 
\bibitem[Berger et al.(2008)]{ber08} Berger, T.~E., Shine, R.~A., Slater, G.~L., et al.\ 2008, \apjl, 676, L89.
\bibitem[Berger et al.(2010)]{ber10} Berger, T.~E., Slater, G., Hurlburt, N., et al.\ 2010, \apj, 716, 1288. 
\bibitem[Berger et al.(2011)]{ber11} Berger, T., Testa, P., Hillier, A., et al.\ 2011, \nat, 472, 197. 
\bibitem[Berger et al.(2017)]{ber17} Berger, T., Hillier, A., \& Liu, W.\ 2017, \apj, 850, 60. 
\bibitem[Bommier et al.(1986)]{bom86} Bommier, V., Leroy, J.~L., \& Sahal-Brechot, S.\ 1986, \aap, 156, 90.
\bibitem[Chae et al.(2008)]{cha08} Chae, J., Ahn, K., Lim, E.-K., et al.\ 2008, \apjl, 689, L73. 
\bibitem[Chandrasekhar(1961)]{cha61} Chandrasekhar, S.\ 1961, International Series of Monographs on Physics, Oxford: Clarendon, 1961.
\bibitem[Chen et al.(2020)]{che20} Chen, P.-F., Xu, A.-A., \& Ding, M.-D.\ 2020, Research in Astronomy and Astrophysics, 20, 166. 
\bibitem[Cheung et al.(2015)]{che15} Cheung, M.~C.~M., Boerner, P., Schrijver, C.~J., et al.\ 2015, \apj, 807, 143. 
\bibitem[de Toma et al.(2008)]{det08} de Toma, G., Casini, R., Burkepile, J.~T., et al.\ 2008, \apjl, 687, L123.
\bibitem[Doschek et al.(1997)]{dos97} Doschek, G.~A., Warren, H.~P., Laming, J.~M., et al.\ 1997, \apjl, 482, L109.
\bibitem[Dud{\'\i}k et al.(2012)]{dud12} Dud{\'\i}k, J., Aulanier, G., Schmieder, B., et al.\ 2012, \apj, 761, 9. 
\bibitem[Franc \& Michel (2004)]{fra04} Franc J. P. \& Michel, J. M. Fundamentals of Cavitation: The dynamics of spherical bubbles CH 3 (Kluwer Academic Publishers, Dordrecht 2004).
\bibitem[Freeland \& Handy(1998)]{fre98} Freeland, S.~L. \& Handy, B.~N.\ 1998, \solphys, 182, 497.
\bibitem[Gonzalez \& Parker(2016)]{gon16} Gonzalez, W. \& Parker, E.\ 2016, Magnetic Reconnection: Concepts and Applications, 427.
\bibitem[Gun{\'a}r et al.(2014)]{gun14} Gun{\'a}r, S., Schwartz, P., Dud{\'\i}k, J., et al.\ 2014, \aap, 567, A123. 
\bibitem[Guo et al.(2021)]{guo21} Guo, Y., Hou, Y., Li, T., et al.\ 2021, \apjl, 911, L9. 
\bibitem[Greisen \& Calabretta(2002)]{gre02} Greisen, E.~W. \& Calabretta, M.~R.\ 2002, \aap, 395, 1061.
\bibitem[Hannah \& Kontar(2012)]{han12} Hannah, I.~G. \& Kontar, E.~P.\ 2012, \aap, 539, A146. 
\bibitem[Hillier et al.(2012)]{hil12} Hillier, A., Berger, T., Isobe, H., et al.\ 2012, \apj, 746, 120. 
\bibitem[Hillier et al.(2012b)]{hil12b} Hillier, A., Hillier, R., \& Tripathi, D.\ 2012, \apj, 761, 106. 
\bibitem[Hillier et al.(2012c)]{hil12c} Hillier, A., Isobe, H., Shibata, K., et al.\ 2012, \apj, 756, 110.
\bibitem[Kaiser et al.(2008)]{kai08} Kaiser, M.~L., Kucera, T.~A., Davila, J.~M., et al.\ 2008, \ssr, 136, 5. 
\bibitem[Labrosse et al.(2011)]{lab11} Labrosse, N., Schmieder, B., Heinzel, P., et al.\ 2011, \aap, 531, A69. 
\bibitem[Landi \& Feldman(2003)]{lan03} Landi, E. \& Feldman, U.\ 2003, \apj, 592, 607.
\bibitem[Landi \& Young(2010)]{lan10} Landi, E. \& Young, P.~R.\ 2010, \apj, 714, 636.
\bibitem[Lemen et al.(2012)]{lem12} Lemen, J.~R., Title, A.~M., Akin, D.~J., et al.\ 2012, \solphys, 275, 17. 
\bibitem[Li et al.(2018)]{li18} Li, D., Shen, Y., Ning, Z., et al.\ 2018, \apj, 863, 192. 
\bibitem[Lin \& Forbes(2000)]{lin00} Lin, J. \& Forbes, T.~G.\ 2000, \jgr, 105, 2375.
\bibitem[Liu et al.(2014)]{liu14} Liu, Z., Xu, J., Gu, B.-Z., et al.\ 2014, Research in Astronomy and Astrophysics, 14, 705-718. 
\bibitem[Mackay et al.(2010)]{mac10} Mackay, D.~H., Karpen, J.~T., Ballester, J.~L., et al.\ 2010, \ssr, 151, 333.
\bibitem[Martin(1998)]{mar98} Martin, S.~F.\ 1998, \solphys, 182, 107. 
\bibitem[Parenti et al.(2005)]{par05} Parenti, S., Lemaire, P., \& Vial, J.-C.\ 2005, \aap, 443, 685. 
\bibitem[Parker(1957)]{par57} Parker, E.~N.\ 1957, \jgr, 62, 509.
\bibitem[Pesnell et al.(2012)]{pes12} Pesnell, W.~D., Thompson, B.~J., \& Chamberlin, P.~C.\ 2012, \solphys, 275, 3
\bibitem[Plesset (1949)]{ple49} Plesset, M. S.\ 1949, Journal of Applied Mechanics, 16, 277–282.
\bibitem[Rayleigh (1917)]{ray17} Rayleigh, Lord.\ 1917, Phil Mag. 34, 94.
\bibitem[Ryutova et al.(2010)]{ryu10} Ryutova, M., Berger, T., Frank, Z., et al.\ 2010, \solphys, 267, 75. 
\bibitem[Samanta et al.(2019)]{sam19} Samanta, T., Tian, H., Yurchyshyn, V., et al.\ 2019, Science, 366, 890. 
\bibitem[Schmieder et al.(2010)]{sch10} Schmieder, B., Chandra, R., Berlicki, A., et al.\ 2010, \aap, 514, A68. 
\bibitem[Shen et al.(2015)]{she15} Shen, Y., Liu, Y., Liu, Y.~D., et al.\ 2015, \apjl, 814, L17. 
\bibitem[Stellmacher \& Wiehr(1973)]{ste73} Stellmacher, G. \& Wiehr, E.\ 1973, \aap, 24, 321
\bibitem[Su et al.(2018)]{su18} Su, Y., Veronig, A.~M., Hannah, I.~G., et al.\ 2018, \apjl, 856, L17. 
\bibitem[Sweet(1958)]{swe58} Sweet, P.~A.\ 1958, Electromagnetic Phenomena in Cosmical Physics, 6, 123.
\bibitem[Thompson \& Wei(2010)]{tho10} Thompson, W.~T. \& Wei, K.\ 2010, \solphys, 261, 215.
\bibitem[Tsiropoula(2000)]{tsi00} Tsiropoula, G.\ 2000, \na, 5, 1. 
\bibitem[Vial \& Engvold(2015)]{via15} Vial, J.-C. \& Engvold, O.\ 2015, Astrophysics and Space Science Library Vol. 415 (Dordrecht: Springer 2015).
\bibitem[Warren \& Brooks(2009)]{war09} Warren, H.~P. \& Brooks, D.~H.\ 2009, \apj, 700, 762.
\bibitem[Wang et al.(2017)]{wan17} Wang, J., Yan, X., Qu, Z., et al.\ 2017, \apj, 839, 128.
\bibitem[Weber et al.(2004)]{web04} Weber, M.~A., Deluca, E.~E., Golub, L., et al.\ 2004, Multi-Wavelength Investigations of Solar Activity, 223, 321.
\bibitem[Wuelser et al.(2004)]{wue04} Wuelser, J.-P., Lemen, J.~R., Tarbell, T.~D., et al.\ 2004, \procspie, 5171, 111. 
\bibitem[Xiang et al.(2016)]{xia16} Xiang, Y.-. yuan ., Liu, Z., \& Jin, Z.-. yu .\ 2016, \na, 49, 8. 
\bibitem[Xue et al.(2020)]{xue20} Xue, Z., Yan, X., Yang, L., et al.\ 2020, \aap, 633, A121. 
\bibitem[Xue et al.(2021b)]{xue21a} Xue, Z.~K., Yan, X.~L., Yang, L.~H., et al.\ 2021b, \apj, 915, 17. 
\bibitem[Xue et al.(2021a)]{xue21} Xue, J.-C., Vial, J.-C., Su, Y., et al.\ 2021a, Research in Astronomy and Astrophysics, 21, 222.
\bibitem[Yan et al.(2015)]{yan15} Yan, X.~L., Xue, Z.~K., Pan, G.~M., et al.\ 2015, \apjs, 219, 17. 
\bibitem[Yan et al.(2015)]{yan15b} Yan, X.-L., Xue, Z.-K., Xiang, Y.-Y., et al.\ 2015, Research in Astronomy and Astrophysics, 15, 1725. 
\bibitem[Yan et al.(2020)]{yan20} Yan, X., Liu, Z., Zhang, J., et al.\ 2020, Science in China E: Technological Sciences, 63, 1656
\bibitem[Yang et al.(2014)]{yang14} Yang, S., Zhang, J., Liu, Z., et al.\ 2014, \apjl, 784, L36. 
\bibitem[Yang et al.(2015)]{yang15b} Yang, Y.-F., Qu, H.-X., Ji, K.-F., et al.\ 2015, Research in Astronomy and Astrophysics, 15, 569.
\bibitem[Yang et al.(2018)]{yang18} Yang, H., Xu, Z., Lim, E.-K., et al.\ 2018, \apj, 857, 115. 
\end{thebibliography}
\end{document}